\documentclass[12pt]{article}
\usepackage{amsmath}
\usepackage{amssymb}
\usepackage{graphicx}
\numberwithin{equation}{section}
\begin{document}
\setcounter{page}{0}
\thispagestyle{empty}
\begin{flushright}
{\small BARI-TH 453/03}
\end{flushright}
\vspace*{2.5cm}
\begin{center}
{\large\bf Dynamics of Ferromagnetic Walls: Gravitational
Properties}
\end{center}
\vspace*{2cm}
\renewcommand{\thefootnote}{\fnsymbol{footnote}}
\begin{center}
{L. Campanelli$^{1,2}$\protect\footnote{Electronic address:
                {\tt campanelli@fe.infn.it}},
P. Cea$^{3,4}$\protect\footnote{Electronic address:
                {\tt Cea@ba.infn.it}}
G.~L. Fogli$^{3,4}$\protect\footnote{Electronic address:
                {\tt Fogli@ba.infn.it}} and
L. Tedesco$^{3,4}$\protect\footnote{Electronic address:
                {\tt luigi.tedesco@ba.infn.it}}\\[0.5cm]
$^1${\em Dipartimento di Fisica,
         Universit\`a di Ferrara, I-44100 Ferrara, Italy}\\[0.2cm]
$^2${\em INFN - Sezione di Ferrara, I-44100 Ferrara, Italy}\\[0.2cm]
$^3${\em Dipartimento di Fisica,
         Universit\`a di Bari, I-70126 Bari, Italy}\\[0.2cm]
$^4${\em INFN - Sezione di Bari, I-70126 Bari, Italy}}
\end{center}
\vspace*{1.0cm}
\vspace*{1.5cm}
\renewcommand{\abstractname}{\normalsize Abstract}
%
%
%
%********************************************************************************************%
%
%
%
\begin{abstract}
We discuss a new mechanism which allows domain walls produced
during the primordial electroweak phase transition. We show that
the effective surface tension of these domain walls can be made
vanishingly small due to a peculiar magnetic condensation induced
by fermion zero modes localized on the wall. We find that in the
perfect gas approximation the domain wall network behaves like a
radiation gas. We consider the recent high-red shift supernova
data and we find that the corresponding Hubble diagram is
compatible with the presence in the Universe of a ideal gas of
ferromagnetic domain walls. We show that our domain wall gas
induces a completely negligible contribution to the large-scale
anisotropy of the microwave background radiation.
\end{abstract}
\vfill
\newpage
%
%
%
%********************************************************************************************%
%
%
%
\renewcommand{\thesection}{\normalsize{\arabic{section}.}}
\section{\normalsize{Introduction}}
\renewcommand{\thesection}{\arabic{section}}
A considerable amount of interest has  emerged in the physics of
topological defects produced during cosmological phase
transitions~\cite{Vilenkin:1994paper}. Indeed, even in a perfectly
homogeneous continuous phase transition, defects will form if the
transition proceeds sufficiently faster than the relaxation time
of the order
parameter~\cite{Kibble:1976sj,Kibble:1980mv,Zurek:1985qw,Zurek:1996sj}.
In such a non-equilibrium transition, the low temperature phase
starts to form, due to quantum fluctuations, simultaneously and
independently in many parts of the system. Subsequently, these
regions grow together to form the broken-symmetry phase. When
different causally disconnected regions meet, the order parameter
does not generally match and a domain structure is formed.
\\
In the standard electroweak phase transition the neutral Higgs
field is the order parameter which is expected to undergo a
continuum phase transition or a
crossover~\cite{Kajantie:1996mn,Fodor:1999at}. In the case in
which the phase transition is induced by the  Higgs sector of the
Standard Model, the defects are domain walls across which the
field flips from one minimum to the other. The defect density is
then related to the domain size and the dynamics of the domain
walls is governed by the surface tension $\sigma$. The existence
of the domain walls, however, is still questionable. It was
pointed out by Zel'dovich, Kobazarev and
Okun~\cite{Zeldovich:1974uw} that the gravitational effects of
just one such wall stretched across the universe would introduce a
large anisotropy into the relic blackbody radiation. For this
reason the existence of such walls was excluded. Quite recently,
however, it has been suggested \cite{Cea:1998ep,Cea:1999jz} that
the effective surface tension of the domain walls can be made
vanishingly small due to a peculiar magnetic condensation induced
by fermion zero modes localized on the wall. As a consequence, the
domain wall acquires a non zero magnetic field perpendicular to
the wall, and it becomes almost invisible as far as gravitational
effects are concerned. In a similar way, even for the bubble
walls, which are relevant in the case of first order phase
transitions, it has been suggested~\cite{Vachaspati:1991nm} that
strong magnetic fields may be produced as a consequence of non
vanishing spatial gradients of the classical value of the Higgs
field.
\\
It is worthwhile to stress that in the realistic case where the
domain wall interacts with the plasma, the magnetic field
penetrates into the plasma over a distance of the order of the
penetration length, which at the epoch of the electroweak phase
transition is about an order of magnitude greater than the wall
thickness. This means that fermions which scatter on the wall feel
an almost constant magnetic field over a spatial region much
greater than the wall thickness.
\\
In this paper we consider thin domain walls eventually produced
during the primordial electroweak phase transition. We focus on
domain walls where the magnetic condensation induced by fermion
zero modes leads to ferromagnetic domain walls (in the following
FDW). Obviously, ferromagnetic domain walls are not topological
stable. However,  FDWs are characterized by the vanishing of the
effective surface tension. As a consequence FDWs are dynamically
stable structures, due to a huge energy barrier against
spontaneous decays, which could have survived until today.
\\
The plan of the paper is as follows. In Section II we consider FDW
in the thin wall approximation and discuss the general properties
of the associated energy-momentum tensor. In Section III we
evaluate the energy momentum tensor contributions due to positive
energy fermions captured by the domain wall. In Section IV we
discuss the equation of state of a domain-wall network in the
ideal gas approximation. In Section V we discuss the gravitational
properties of the ideal gas of FDWs. Section VI is devoted to the
analysis of the temperature anisotropy on the Cosmic Microwave
Background Radiation induced by the ferromagnetic domain wall
network. Finally, we summarize our results in Section VII. Some
technical details are relegated in the Appendix.
%
%********************************************************************************************%
%
%
%
\renewcommand{\thesection}{\normalsize{\arabic{section}.}}
\section{\normalsize{The energy-momentum tensor of a FDW }}
\renewcommand{\thesection}{\arabic{section}}
Topologically stable kinks are ensured when the vacuum manifold
${\cal M}$, the space of all accessible vacua of the theory, is
disconnected, {\it i.e.} $\pi_{0}({\cal M})$ is nontrivial.
\\
We shall consider a simplified model in which the kink is a
infinitely static domain wall in the $(x,y)$ plane in a flat
space-time. That is we assume that the vacuum manifold consists of
just two disconnected components and restrict ourselves to one
spatial dimension. In our model the scalar sector that give rise
to a planar wall is given by a real scalar field with density
Lagrangian,
\begin{equation}
{\cal L}_{\phi} = \frac{1}{2} \, \partial^{\mu} {\phi} \,
\partial_{\mu} {\phi} \, - V(|\phi|,T).
\end{equation}
Lagrangian (2.1) has an exact $Z_{2}$ symmetry corresponding to
the discrete transformation $\phi \rightarrow -\phi$. At zero
temperature the $Z_{2}$ symmetry is spontaneously broken, with $
\phi$ acquiring a vacuum expectation of order $v$. In the tree
approximation, the set of vacuum states is $\langle \, \phi \,
\rangle^2 = v^2$ so that $\pi_{0}({\cal M}) \neq 0$. Then, if we
apply boundary conditions that the vacuum at $- \infty$ lie in
$-v$ and that at $+ \infty$ lie in $+v$, by continuity there must
exists a region in which the scalar field is out of the vacuum.
This region is a domain wall. If we take the potential at zero
temperature as $V(\phi) = \frac{\lambda}{4} ({\phi}^{2} -
v^{2})^{2}$, it is easy to see that the classical equation of
motion admits the solution describing the transition layer between
two regions with different values of $\langle \, \phi \, \rangle$,
\begin{equation}
\phi(z) = v \, \tanh (z / \Delta),
\end{equation}
where $\Delta = \sqrt{2} \lambda^{-1/2} \, v^{-1}$ is the
thickness of the wall.
\\
The energy-momentum tensor for the $\phi$-kink (2.2) is,
\begin{equation}
T_{(\phi) \; \nu}^ {\mu} = \alpha(z) \, \sigma_{\phi} \; diag \,
(1,1,1,0),
\end{equation}
\begin{equation}
\alpha(z) = \frac{3}{4\Delta}\,[\, \cosh{(z\slash{\Delta})}]^{-4},
\end{equation}
where
\begin{equation}
\sigma_{\phi} = \int_{- \infty}^{+ \infty}dz \, T_{(\phi) \; 0}^
{0} = \frac{2\sqrt{2}}{3} \lambda^{1/2} v^{3}
\end{equation}
is the surface energy density associated with the scalar field. In
the thin wall approximation, {\it i.e.} $\Delta \rightarrow 0$,
$\alpha(z)$ reduces to a $\delta$-function. Unless the symmetry
breaking scale $v$ is very small, surface density energy of the
kink is extremely large and implies that cosmological domain walls
would have an enormous impact on the homogeneity of the Universe.
A stringent constraint on the wall tension $ \sigma_{\phi}$ for a
$Z_2$-wall can be derived from the isotropy of the microwave
background. If the interaction of walls with matter is negligible,
then there will be a few walls stretching across the present
horizon. They introduce a fluctuation in the temperature of the
microwave background of order
$\delta T / T \sim G_N \, \sigma_{\phi} \, t_0$,
where $t_0$ is the present time. Observations constrain $\delta T
/ T \, < \, 10^{-5}$, and thus models predicting topologically
stable domain walls with $v \, > \, 1 \;  MeV$ should be ruled
out.
\\
On the other hand, we are interested in domain walls formed during
the electroweak phase transition by the Kibble-Zurek
mechanism~\cite{Kibble:1976sj,Kibble:1980mv,Zurek:1985qw,Zurek:1996sj}.
Even these domain walls are approximately accounted for by
Eq.~(2.2), so that the energy-momentum tensor is given by
Eq.~(2.3). Since $v \, \sim \, 10^2 \;  GeV$, it is widely
believed that there must be some provision for the elimination of
these defects which indeed  are not topological stable. However,
quite recently it has been suggested that the effective surface
tension of the domain wall can be made vanishingly small due to a
peculiar magnetic condensation induced by fermion zero modes
localized on the wall. As a consequence the domain wall acquires a
non zero magnetic field perpendicular to the wall. This mechanism
is able to make these defects almost dynamical stable. Indeed,
FDWs  are protected against decay by a huge energy barrier.
\\
In general, for a FWD the magnetic field vanishes in the regions
where the scalar condensate is constant. So that, the magnetic
field can be different from zero only in the regions where the
scalar condensate varies, $i.e.$ in a region of the order of the
wall thickness. However, it is worthwhile to stress that in the
realistic case where the domain wall interacts with the plasma,
the magnetic field penetrates into the plasma over a distance of
the order of the penetration length which at the time of
electroweak phase transition is about an order of magnitude
greater than the wall thickness. This means that fermions
interacting with the wall feel an almost constant magnetic field
over a spatial region much greater than the wall thickness. So
that as concern the interaction of fermions with FWDs, we can
assume that the magnetic field is almost constant.
\\
For completeness, let us briefly review the generation of
ferromagnetic domain walls at the electroweak phase transition
\cite{Cea:1998ep}. After the electroweak spontaneous symmetry
breaking the unique long range gauge field is the electromagnetic
field $A_{\mu}$. In our scheme the real scalar field is the
physical Higgs field, so that $A_{\mu}$ does not couple directly
to $\phi$. The electromagnetic Lagrangian is:
\begin{equation}
{\cal L}_{em} = - \frac{1}{4} F_{\mu \nu} \, F^{\mu \nu}.
\end{equation}
We are interested in the case of an external magnetic field $B$
(localized on the wall at $z=0$) directed along the $z$-direction,
\begin{equation}
A_{\mu} = (0,0,-Bx,0),
\end{equation}
where without loss of generality we may consider $B>0$. Assuming
that the magnetic field $B$ is localized on the wall, we can
write,
\begin{equation}
T_{(B) \; \nu}^ {\mu}(z) = \beta(z) \, \sigma_{B} \; diag \,
(1,-1,-1,0),
\end{equation}
where
\begin{equation}
\sigma_{B} = \int_{- \infty}^{+ \infty}\!dz \, T_{(B) \; 0}^ {0} =
\Delta \frac{B^{2}}{2}
\end{equation}
is the surface energy density associated with the magnetic field
if $B$ is sizable over a distance $z \sim \Delta$, and $ \beta(z)$
is a localization function of $B$ on the wall. In the thin wall
approximation we expect that $\beta(z)$ reduces to a
$\delta$-function.
\\
Next we consider a four dimensional massless Dirac fermion coupled
to the scalar field through the Yukawa coupling, and in presence
of the homogeneous background magnetic field (2.7). The Dirac
fermion $\psi$ is described by the Lagrangian:
\begin{equation}
{\cal L}_{\psi}= {\bar{\psi}}(i \, \slash\!\!\!\partial - e \;
\slash\!\!\!\!\!\:\!A - g_{Y} \, \phi) \, \psi
\end{equation}
where we consider $e>0$. In presence of the kink (2.2) the Dirac
equation,
\begin{equation}
(i \, \slash\!\!\!\partial - e \; \slash\!\!\!\!\!\:\!A - g_{Y} \,
\phi ) \, \psi = 0,
\end{equation}
admits solutions localized on the wall~\cite{Cea:1998ep}. Using
the following representation of the Dirac algebra,
\begin{equation}
\alpha = 0,1,2 \, : \;
               \gamma \, ^{\alpha} = \left(
                          \begin{array}{cc}
                          \stackrel{\sim}{\gamma} \!\!\, ^\alpha & 0 \\
                          0 & -\stackrel{\sim}{\gamma} \!\!\, ^\alpha
                          \end{array}
                      \right) \! , \;\;\;\;
                \gamma \, ^{3} = \left(
                          \begin{array}{cc}
                          0 & i \\
                          i & 0
                          \end{array}
                      \right) \!,
\end{equation}
with
\begin{equation}
\stackrel{\sim}{\gamma} \!\!\, ^0 = \sigma_{3}, \;\;
\stackrel{\sim}{\gamma} \!\!\, ^1 = i \,\sigma_{1}, \;\;
\stackrel{\sim}{\gamma} \!\!\, ^2 = i \,\sigma_{2},
\end{equation}
where $\sigma_{i}$ are the usual Pauli matrices, we find for the
localized modes \cite{Cea:1998ep},
\begin{equation}
\psi (x,y,z,t) = \frac{1}{\sqrt{2}} \, \omega(z)
                          \left( \begin{array}{c}
                          \rho(x,y,t) \\
                          \rho(x,y,t)
                          \end{array} \right),
\end{equation}
\begin{equation}
\omega(z) = N \, [\, \cosh{(z\slash{\Delta})}]^{-\xi \Delta},
\end{equation}
where $\xi = g_{Y} \, v$ is the mass which the fermion acquires in
the broken phase, $\rho(x,y,t)$ is a Pauli spinor that satisfies
the $(2+1)$-dimensional massless Dirac equation in a constant
external magnetic field, and $N$ is a normalization constant given
by
\begin{equation}
N = \left[ \: \int_{- \infty}^{+ \infty} \!\! dz\,
     \omega^{2}(z) \right]^{-1/2} \! = \:
      \left[ \Delta \, \mbox{B}(\xi \, \Delta ,1/2) \right]^{-1/2} ,
\end{equation}
 B$(x \, ,y)$ being the Euler Beta function.
\\
Let us consider the energy-momentum tensor of the four dimensional
massless Dirac fermion, $T^{\mu}_{(\psi) \, \nu}$. It is
straightforward to show that the non-zero components are given by:
\begin{equation}
T^{\alpha}_{(\psi) \, \beta} = i \, \omega^2(z) \rho^{\dag}
 \stackrel{\sim}{\gamma}\!\!\, ^0
 \stackrel{\sim}{\gamma}
 \!\!\, ^\alpha  \, \partial_{\beta} \rho ,
\end{equation}
\begin{equation}
T^{\alpha}_{(\psi) \, 3}= -i \, g_Y \, \phi(z) \omega^2(z)
 \rho^{\dag}
 \stackrel{\sim}{\gamma}\!\!\,^0
 \stackrel{\sim}{\gamma}
 \!\!\, ^\alpha \rho.
\end{equation}
In the thin-wall approximation $ \omega^2(z)$ reduces to a
$\delta$-function so that  $T^{\alpha}_{(\psi) \, 3}$ goes to
zero. Therefore the tensor reduces to:
\begin{equation}
\label{eq2.20} T_{(\psi) \; \nu}^{\,\mu} (x,y,z) = \delta(z) \,
       \left(
          \begin{array}{cc}
           T_{(\rho) \; \beta}^{\,\alpha} (x,y) \;\; & \;\; 0 \\
           \; \, \; \\
           0 \;\;                                   & \;\; 0
          \end{array}
       \right),
\end{equation}
where $T_{(\rho)\beta}^{\,\alpha} =
     i \rho^{\dagger}\stackrel{\sim}{\gamma} \!\!^0
    \stackrel{\sim}{\gamma} \!\!^{\alpha}
    \partial_{\beta} \rho$
is the energy-momentum tensor for a $(2+1)$-dimensional massless
Dirac fermion coupled to the magnetic field.
\\
The resulting energy-momentum tensor of the wall is given by
\begin{equation}
T_{(w) \; \nu}^ {\mu} = T_{(\phi) \; \nu}^ {\mu} + T_{(B) \; \nu}^
{\mu} + T_{(\psi) \; \nu}^ {\mu}.
\end{equation}
>From Eq.~(2.19) we see that the $(3+1)$-dimensional tensor
$T^{\mu}_{(\psi) \, \nu}$ is obtained by evaluating
$T_{(\rho)\beta}^{\,\alpha}$ which corresponds to
$(2+1)$-dimensional Dirac fermions localized on the wall.
\\
It is known since long time that in $(2+1)$-dimensional QED in
presence of un external homogeneous background magnetic field with
massless Dirac fermions it is energetically favorable the
spontaneous generation of a negative mass term~\cite{Cea:1985}. So
we are lead to consider the following effective
$(2+1)$-dimensional Dirac equation:
\begin{equation}
(i \stackrel{\sim}{\gamma}\!\!^{\alpha}
\partial_{\alpha} - e \stackrel{\sim}{\gamma}\!\!^{\alpha}
A_{\,\alpha} - m )\, \rho(x,y,t) = 0.
\end{equation}
Taking $\rho(x,y,t) = e^{-iEt}\,\rho(x,y)$ we get~\cite{Cea:1985}
\begin{equation}
E=+E_n \, : \;\;\; \rho_{n,p}^{(+)} =
                    \left( \frac{E_n + m}{2E_n} \right)^{1/2}
                    \frac{e^{ipy}}{\sqrt{2\pi}}\;
                    e^{-\zeta^{2} /2}\,
                    \left(
          \begin{array}{c}
            N_{n} \, H_{n}(\zeta) \\
           \, \\
           -\frac{(E_{n}^{2} - m^{2})^{1/2}}{E_n + m} \, N_{n-1} \,
                                                         H_{n-1}(\zeta)
          \end{array}
                    \right) \! ,
\end{equation}
\begin{equation}
E=-E_n \, : \;\;\; \rho_{n,p}^{(-)} =
                    \left( \frac{E_n - m}{2E_n} \right)^{1/2}
                    \frac{e^{ipy}}{\sqrt{2\pi}}\;
                    e^{-\zeta^{2} /2}\,
                    \left(
          \begin{array}{c}
            N_{n} \, H_{n}(\zeta) \\
           \, \\
           \frac{(E_{n}^{2} - m^{2})^{1/2}}{E_n + m} \, N_{n-1} \,
                                                         H_{n-1}(\zeta)
          \end{array}
                    \right) \! ,
\end{equation}
\begin{equation}
E=E_0=m \, : \;\;\; \rho_{0,p} =
                    N_{0}\,\frac{e^{ipy}}{\sqrt{2\pi}}\;
                    e^{-\zeta^{2} /2}\,
                    \left(
          \begin{array}{c}
            1 \\
            0
          \end{array}
                    \right) \! ,
\end{equation}
where
\begin{eqnarray}
E_{n} \!\! & = & \!\! \sqrt{2 n eB + m^2} \; , \\
\zeta \!\! & = & \!\! \sqrt{eB} \left(x-\frac{p}{eB} \right), \\
N_n \!\! & = & \!\!
      \left( \frac{eB}{\pi} \right)^{1/4} \frac{1}{\sqrt{2^{n} \, n!}} \,,
\end{eqnarray}
with $n = 1,2,...$, and $H_{n}(x)$ are Hermite polynomials. The
normalization is
\begin{equation}
\int\!\!\ d^{2}x \; \rho_{n,p}^{(\pm)\;\dag} ({\bf x}) \,
\rho_{n',p'}^{(\pm)} ({\bf x}) = \delta(p-p\,') \, \delta_{nn'} \,
.
\end{equation}
It is convenient to expand the fermion field operator $\rho$ in
terms of $\rho_{n,p}^{(+)}$ and $ \rho_{n,p}^{(-)}$. In other
words, we adopt the so-called Furry's representation.  For
negative mass term we have:
\begin{equation}
\rho \; = \;
         \sum_{n=0}^{\infty} \int_{- \infty}^{+ \infty}\!\!dp \;
          a_{n \, p} \, \rho_{n,p}^{(+)} \; + \;
          \sum_{n=1}^{\infty} \int_{- \infty}^{+ \infty}\!\!dp \;
          b_{n \, p}^{\dagger} \, \rho_{n,p}^{(-)} \; .
\end{equation}
We observe that in the case of negative mass the positive
solutions have eigenvalues $+E_n$ with $n\geq 1$ and the negative
ones $-E_n$ with $n\geq 0$. In the expansion of $\rho$ we
associate particle operators to positive energy solutions and
antiparticle operators to negative energy solutions. These
operators satisfy the standard anticommutation relations,
\begin{equation}
\{a_{n \, p} \; , a_{n' \, p'}^{\dagger}\} = \{b_{n \, p} \;,
b_{n' \, p'}^{\dagger}\} = \delta_{n \, n'} \, \delta(p-p \, '),
\end{equation}
others anticommutators being zero. The energy-momentum tensor
operator is
\begin{equation}
T_{(\rho) \; \beta}^{\,\alpha} =
     i \,  \rho^{\dagger} \, \stackrel{\sim}{\gamma} \!\!\, ^0
    \stackrel{\sim}{\gamma} \!\!\, ^{\alpha} \,
    \partial_{\beta}\; \rho.
\end{equation}
We need  the surface density of the vacuum expectation value of
energy-momentum tensor operator defined as:
\begin{equation}
\langle \, T_{(\rho) \; \beta}^{\,\alpha} \, \rangle = \frac{1}{A}
\int d^2 \, x \, \frac {\langle 0 | \, T_{(\rho) \;
\beta}^{\,\alpha} \, | 0 \rangle} {\langle 0 | 0 \rangle} \, ,
\end{equation}
where $A$ is the area of the wall and $|0 \rangle$ is the fermion
vacuum. In the Furry's representation $\langle \, T_{(\rho) \;
\beta}^{\,\alpha} \, \rangle$  reads
\begin{equation}
\langle \, T_{(\rho) \; \beta}^{\,\alpha} \, \rangle =
      \frac{1}{A} \int \!d^{2}x \,
      \sum_{n=0}^{\infty} \, \int_{- \infty}^{+ \infty}\!\!dp \;
        \rho_{n,p}^{(-) \, \dagger} \; i
            \stackrel{\sim}{\gamma} \!\!\, ^0
            \stackrel{\sim}{\gamma} \!\!\, ^{\alpha} \,
            \partial_{\beta} \;
        \rho_{n,p}^{(-)} \; .
\end{equation}
for the negative mass theory. Inserting Eqs. (2.23) and (2.24)
into Eq. (2.33), it is straightforward to obtain
\begin{eqnarray}
    \sigma_{\rho} \! & \equiv & \!
            \langle \, T_{(\rho) \; 0}^{\,0} \, \rangle \; = \;
            -\frac{eB}{2\pi} \sum_{n=0}^{\infty} E_{n}, \\
     \, \nonumber \\
\label{eq2.35}
     - p_{\rho} \! & \equiv & \!
             \langle \, T_{(\rho) \; 1}^{\,1} \, \rangle \; = \;
             \langle \, T_{\rho \; 2}^{\,2} \, \rangle \; = \;
             \frac{{(eB)}^{2}}{2\pi} \sum_{n=0}^{\infty}
             \frac{n}{E_{n}} \, ,
\end{eqnarray}
while all other $\langle \, T_{(\rho) \; \beta}^{\, \alpha} \,
\rangle$ are equal to zero. Therefore, the energy-momentum tensor
is:
\begin{equation}
\langle \, T_{(\rho) \; \beta}^{\, \alpha} \, \rangle= diag \,
(\sigma_{\rho},-p_{\rho},-p_{\rho}),
\end{equation}
where $\sigma_{\rho}\,$ and $\, p_{\rho}$ are the surface density
energy and pressure of the fermionic condensate, respectively. It
easy to show that the following relation
\begin{equation}
p_{\rho}= B \frac {\partial \sigma_{\rho}} {\partial B} -
\sigma_{\rho}
\end{equation}
holds. We shall consider the ``effective" energy-momentum tensor
defined as:
\begin{equation}
\langle \, T_{(\rho) \; \beta}^ {\alpha} \, \rangle_{eff} \, = \,
\langle \, T_{(\rho) \; \beta}^ {\alpha} \, \rangle_{B} \; -
\langle \, T_{(\rho) \; \beta}^ {\alpha} \, \rangle_{B=0} \, .
\end{equation}
By using the integral representation
\begin{equation}
\sqrt{a} = - \int_{0}^{\infty}\!\frac {ds} {\sqrt{\pi s}} \frac
{d} {ds} \, e^{-as},
\end{equation}
introducing the dimensionless variable $\, \eta = eB/{m^2} \,$ and
the function
\begin{equation}
g(\eta) = \int_{0}^{\infty}\!\frac {ds} {\sqrt{\pi s}}
\frac{d}{ds} \, e^{-s\slash{\eta}} \left[ \frac{1}{1 - e^{-2s}} -
\frac{1}{2s} \right] \! ,
\end{equation}
we cast Eq. (2.38) into
\begin{eqnarray}
  \sigma_{\rho}^{eff} \! & \equiv & \!
       \langle \, T_{(\rho) \; 0}^{0} \, \rangle_{eff} \; = \;
       \delta(z) \,\frac{(eB)^{3/2}}{2 \pi} \, g(\eta), \\
       \label{eq2.42}
 - p_{\rho}^{eff} \! & \equiv & \!
       \langle \, T_{(\rho) \; 1}^{1} \, \rangle_{eff} \; = \;
       \langle \, T_{(\rho) \; 2}^{2} \, \rangle_{eff} \; = \;
      - \delta(z) \,\frac{(eB)^{3/2}}{2 \pi} \, \left[\frac {1} {2} \, g(\eta) +
       \eta \, \frac {\partial g (\eta)} {\partial \eta} \right] \!
       .
\end{eqnarray}
Taking into account Eqs. (2.20), (2.41), (2.42), the final
expression of the effective energy-momentum tensor of a thin FDW
is:
\begin{equation}
\langle \, T_{(w) \; \nu}^ {\mu} \, \rangle = \delta(z) \, diag
\left(\,\sigma_{w} \, , \, -p_{w} \, , \, -p_{w} \, , 0\right),
\end{equation}
where
\begin{eqnarray}
\sigma_{w} \!\! & = & \!\! \sigma_{\phi} + \sigma_{B} +
\sigma_{\rho}^{eff}, \\
\label{eq2.45} p_{w} \!\! & = & \!\! -\sigma_{\phi} + \sigma_{B} +
p_{\rho}^{eff}.
\end{eqnarray}
We are interested on the value of $\langle T^{\mu}_{(w) \, \nu
}\rangle$ for the magnetic field $B^*$ that minimizes the energy
density $\sigma_w$. As shown in Ref.~\cite{Cea:1998ep} the
magnetic field $B^*$  can be approximated by:
\begin{equation}
B^*=\frac {e |m|^2} {4 \pi} \frac {1} {|m| \Delta + \frac {e^2}
{12 \pi}} \; .
\end{equation}
Moreover the stability condition
\begin{equation}
\sigma_w=0
\end{equation}
gives in turns the value  of $|m^*|$. Equation~(2.47) assures that
our ferromagnetic domain walls have a vanishing effective surface
tension. Moreover, it is remarkable that Eq.~(2.47) implies:
\begin{equation}
p_w(B^*,|m|^*)=0 \; .
\end{equation}
Note that Eq.~(2.48) is an exact consequence of the stability
condition Eq.~(2.47) and does not rely on the approximate
expression for $B^*$ given by Eq.~(2.46). By using the standard
values $v \simeq 250 \!$ GeV, $ \alpha_{QED} = e^2 / 4 \pi \simeq
1/137$ and assuming $\lambda \simeq 0.5$ we obtain from
Eqs.~(2.46) and (2.47) the following value of the magnetic
field~\cite{Cea:1998ep},
\begin{equation}
B^* \simeq 5.0 \; 10^{24} \, \mbox{Gauss}.
\end{equation}
Interestingly enough, such a value of the magnetic field at the
primordial  electroweak phase transition could be relevant for the
generation of the primordial magnetic field~\cite{Dolgov01}.
\\
It is worthwhile to stress that Eqs.~(2.47) and (2.48) imply that
our ferromagnetic domain walls are almost invisible as concern the
gravitational effects. However, as already discussed, in the
realistic case where the domain wall interacts with the plasma,
the magnetic field penetrates into the plasma over a distance of
the order of the penetration length $\lambda$ which is about an
order of magnitude greater than $\Delta $. It follows that the
above estimate  of the induced magnetic field at the electroweak
scale $ B^* $ is reduced by a factor $\sqrt{\frac {\Delta}
{\lambda}} \, \sim \, 0.3$ which is still of cosmological
interest~\cite{Cea:2002iy}. Moreover, on the ferromagnetic domain
walls there are also  positive energy states. As a consequence
fermions incident on the wall with an energy equal to the empty
states  can be trapped on the wall giving a non trivial
contribution to the energy-momentum tensor. Thus, in order to
investigate the gravitational properties of the ferromagnetic
domain walls it is important to take care of these contributions
to the energy-momentum tensor.
%
%
%
%********************************************************************************************%
%
%
%
\renewcommand{\thesection}{\normalsize{\arabic{section}.}}
\section{\normalsize{The energy-momentum tensor of a massive  FDW }}
\renewcommand{\thesection}{\arabic{section}}
On the wall there are also available positive energy bounded
states. Therefore when incident fermions have the same energy as
the allowed bounded states on the wall, they could be captured. In
this case, the fermions bounded to the ferromagnetic domain wall
do contribute to the energy-momentum tensor of the wall. In the
following we shall refer to FDW bounded with positive-energy
fermions as massive FDW. In this Section we calculate the
energy-momentum tensor of massive FWDs.
\\
To discuss the localized solutions with positive energy we rewrite
Eq.~(2.11) as:
\begin{equation}
(i \, \slash\!\!\!\partial - e \; \slash\!\!\!\!\!\:\!A - \xi \,
g) \, \psi(x,y,z,t)=0,
\end{equation}
where $\xi=g_Y \,v$, $\, g(z)= \tanh (\frac {z} {\Delta})$ and $
\psi(x,y,z,t)$ is the wave function of the fermion eventually
captured by the FDW and having positive energy corresponding to a
allowed bound state on the wall. For reader convenience we
relegate in the Appendix the details of our calculations. As shown
in the Appendix, the wave functions of the localized fermionic
modes are given by:
\begin{eqnarray}
\psi^1_{+e} & = & N \, A_n \, e^{- \frac {\xi^2} {2}+ i \, p_y \,
y - i\, E\, t } (\cosh \frac {z} {\Delta})^{- \xi \, \Delta} \,
[H_n \, E \, u^1_+ + 2\, n \, \sqrt{|eB|} \, H_{n-1} \, u^2_{+} \,
],
\\
\label{eq3.3} \psi^2_{+e} & = & N \, A_{n-1} \, e^{- \frac {\xi^2}
{2}+ i \, p_y \, y - i\, E\, t } (\cosh \frac {z} {\Delta})^{- \xi
\, \Delta} \, [H_{n-1} \, E \, u^2_+ + \sqrt{|eB|} \, H_{n} \,
u^1_{+} \, ],
\\
\label{eq3.4} \psi^1_{-e} & = & N \, A_{n-1} \, e^{- \frac {\xi^2}
{2}+ i \, p_y \, y - i\, E\, t } (\cosh \frac {z} {\Delta})^{- \xi
\, \Delta} \, [H_{n-1} \, E \, u^1_+ - \sqrt{|eB|} \, H_{n} \,
u^2_{+} \, ],
\\
\label{eq3.5} \psi^2_{-e} & = & N \, A_n \, e^{- \frac {\xi^2}
{2}+ i \, p_y \, y - i\, E\, t } (\cosh \frac {z} {\Delta})^{- \xi
\, \Delta} \, [H_n \, E \, u^2_+ -2 \, n \,  \sqrt{|eB|} \,
H_{n-1} \, u^1_{+} \, ],
\end{eqnarray}
where $N$ is the normalization constant evaluated in the Appendix,
and
\begin{equation}
E_n  =  \sqrt{2 \, n \, |e|\, B}, \, \, \,\, \, \,  \, n=1,2,...
\end{equation}
The wave functions (3.2), (3.3) correspond to fermions, while
(3.4), (3.5) to antifermions. Note that fermions have spin
parallel to the magnetic field, while antifermions have spin
antiparallel to the magnetic field.
\\
In a previous paper~\cite{Campanelli02} we investigated the
scattering of fermions off walls in presence of a constant
magnetic field \footnote{Similar calculations have been presented
in Ref. \cite{AYALA}.}. In particular, we discussed fermion states
corresponding to solutions of Dirac equation (2.11) which are
localized on the domain wall. We found that there is total
reflection for fermions with parallel and antiparallel spin at
energies $E_n^2 \,-\, \xi^2 =  2 \, n \, |e|\, B$ and $E_n^2 \,-\,
\xi^2 = 2 \, (n+1) \, |e|\, B$, respectively. Note that the
difference is due to the fermion mass term $\xi$ which, indeed,
vanishes on the wall where the system is in the symmetric phase.
>From the physical point of view, we see that fermions with
asymptotically positive kinetic energy equal to $\sqrt{2 \, n \,
|e|\, B}$ for parallel spin, and $\sqrt{2 \, (n+1) \, |e|\, B}$
for antiparallel spin, can be indeed trapped on the domain wall.
As a consequence, we see that ferromagnetic domain walls immersed
into the primordial plasma are able to capture fermions by filling
the Landau levels up to $n \, = \, N_{max}$. Obviously, $N_{max}$
depends on the plasma temperature $T$, for  we must have $E_n \;
\lesssim \; T$. In our case, the temperature at the electroweak
phase transition is of the order of $10^2 \; GeV$.  Taking into
account Eqs.~(2.49) and (3.6) we get $N_{max} \; \backsimeq \; 1$.
\\
Let us indicate with $| w \rangle$ the generic massive FDW, namely
a ferromagnetic domain wall filled with  fermions ($+e$) and
antifermions ($-e$), with ``spin up'' ($s=1$) and ``spin down''
($s=2$) respectively, up to  Landau level  $N_{max}$ . Thus, the
expectation value of the energy-momentum tensor on the state $| w
\rangle$ is:
\begin{equation}
\langle T^{\mu}_{(\Psi)\, \nu} \rangle =
  \frac {1} {A} \int d^2x \,
  \frac {\langle \, w | T^{\mu}_{(\psi)\, \nu} | w \, \rangle}
  {\langle w | w \rangle} \, ,
\end{equation}
where
\begin{equation}
T^{\mu}_{(\Psi) \, \nu}= i \, {\bar {\Psi}} \gamma^{\mu}
\partial_{\nu} \Psi \; .
\end{equation}
To evaluate $\langle T^{\mu}_{(\Psi) \, \nu} \rangle$  we expand
the fermionic operator $\Psi$ in terms of the states (3.2)-(3.5)
as follows
\begin{equation}
\Psi=\sum_{n_s} \sum_s \int dp_y \, [\,a(u_s,p_y,s)\,
\psi_{+e}^{s,n_s,p_y} + b^{\dag}(u_s,p_y,s)\,
\psi^{s,n_s,p_y}_{-e}\, ].
\end{equation}
In the thin wall approximation we find,
\begin{equation}
\langle T^{\mu}_{(\Psi)\, \nu} (z) \rangle = \delta (z) \,
\sigma_{\Psi} \, diag (1,-1/2,-1/2,0),
\end{equation}
where
\begin{equation}
\sigma_{\Psi}= \int d\, z\, \langle T^0_{(\Psi)\, 0}(z) \rangle =
\frac {2 \, |e \, B^*|^{3/2}} {\pi} \, \sum_{n=1}^{N_{max}}
\sqrt{n} \; \backsimeq \; \frac {2 \, |e \, B^*|^{3/2}} {\pi} \, .
\end{equation}
Taking into account the value of $B^*$  given in Eq.~(2.49), we
see that Eq.~(3.11) implies that $\sigma_{\Psi}$ is comparable to
$\sigma_{\phi}$ given by Eq.~(2.5), so that massive FDWs could
display important gravitational effects which, eventually, could
be in contrast with observations. Indeed, the cosmological
electroweak phase transition should lead to a network of domain
walls. However, solving for the cosmic evolution of a domain-wall
network is quite involved. Nevertheless some essential features
can studied by considering the dynamic of the domain-wall network
in the ideal gas approximation~\cite{KOLB}.
%
%
%
%********************************************************************************************%
%
%
%
\renewcommand{\thesection}{\normalsize{\arabic{section}.}}
\section{\normalsize{Ideal gas of FDWs}}
\renewcommand{\thesection}{\arabic{section}}
In this Section we shall follow quite closely the analysis of
Ref.~\cite{KOLB} to calculate the equation of state for a ideal
gas of massive ferromagnetic domain walls.
\\
Let us consider a perfect gas of walls moving with velocity $v$ in
a box of volume $V \; \gg \; \xi$, where $\xi$ is a typical
distance between walls. The perfect gas approximation amounts to
neglect any dissipative effects due to the interaction of the
walls. We first consider a planar wall in the $x-y$ plane. Because
of the symmetry, the energy-momentum tensor will be a function
only of $z$. Following Ref.~\cite{KOLB} we find that the average
energy-momentum tensor of the wall gas is:
\begin{equation}
\label{eq4.1}
\langle T^{\mu \nu}_{(w)} \rangle \sim
  \frac{1}{\langle L \rangle} \int \!dz \;
    T^{\mu \nu}_{(w)} (z) \equiv
    \frac{1}{\langle L \rangle} W^{\mu \nu}.
\end{equation}
where $T^{\mu \nu}_{(w)}$ is given by Eqs.~(3.10) and (3.11) for
massive FDWs, and $\langle L \rangle$ is the average wall
separation. If the walls are moving with average velocity $v$ in
the $+ \, \hat{z}$ direction, we can obtain the energy-momentum
tensor by boosting $W^{\mu \nu}$ in $z$-direction. Repeating the
procedure for walls moving in the $- \, \hat{z}$,  $\pm \,
\hat{x}$, $\pm \, \hat{y}$ directions and averaging over the
walls, we see that the average energy-momentum tensor for an ideal
gas of massive FDWs is diagonal. Moreover, we have:
\begin{eqnarray}
\label{eq4.2}
\rho_{w} &\!\! \equiv \!\!&  \langle T^{00}_{(w)} \rangle =
  \frac{\sigma_{w}}{\langle L \rangle} \gamma^{2} , \\
\label{eq4.3}
p_{w} &\!\! \equiv \!\!& \langle T^{11}_{(w)} \rangle =
     \langle T^{22}_{(w)} \rangle =
     \langle T^{33}_{(w)} \rangle =
  \frac{1}{3} \, \rho_{w} \, ,
\end{eqnarray}
where $\gamma = \left(1-v^2\right)^{-1/2}$ is the Lorentz factor
and $v$ is the mean velocity of a wall. In Eq.~(4.2) $\sigma_{w}$,
the surface energy density for a massive FDW, is given by
Eq.~(3.11). From Eq.~(4.3) we see that the equation of state for a
ideal gas of massive FDWs does not depend on the wall velocity and
coincides with the equation of state of a radiation gas.

It is clear that in the more realistic case in which the walls
interact with the primordial plasma the equation of state for a
gas of massive FDWs is expected to be more stiff than that of a
radiation gas. Writing the pressure as $p_w = \alpha \rho_w$ we
expect that $\alpha < 1/3$ in the primordial Universe (say before
the time of structure formation). On the other hand, at the
present time the dissipative effects due to the interaction of the
walls can be safely neglected and then the perfect gas
approximation holds.

In the following we shall work in the ideal case of
non-interacting walls since we will concentrate on physical
process occurred at relatively recent times (i.e. after the
beginning of structure formation).
%
%
%
%********************************************************************************************%
%
%
%
\renewcommand{\thesection}{\normalsize{\arabic{section}.}}
\section{\normalsize{Large scale  gravitational properties}}
\renewcommand{\thesection}{\arabic{section}}
In a recent paper we investigated the gravitational properties of
thin planar massive ferromagnetic domain walls in the weak field
approximation~\cite{Campanelli:2003nf}. In order to understand the
cosmological implications of FDWs we must consider the standard
model of the Universe which is based upon the
Friedmann-Robertson-Walker cosmological model~\cite{KOLB}. The
Friedmann-Lemaitre equations for the expansions parameter $R$ are:
\begin{equation}
\frac{{\stackrel{\cdot}{R}}{\,^2}}{R^2} + \frac{k}{R^2} =
\frac{8\pi G_N}{3}\,\rho_{tot}  \, + \, \frac{\Lambda}{3},
\end{equation}
\begin{equation}
2 \frac{\stackrel{\cdot \cdot}{R}}{R} +
\frac{{\stackrel{\cdot}{R}}{\,^2}}{R^2} + \frac{k}{R^2} = -8\pi
G_N \,p_{tot} \; + \; \Lambda \; ,
\end{equation}
where $\stackrel{\cdot}{R} = dR/dt$ is the derivative respect to
the cosmic time $t$, $k$ is the curvature constant ( $k$ equals to
$-1,\,0,\,+1$ for a Universe which is respectively open, flat, and
closed), and $\Lambda$ is the cosmological constant.
Equation~(5.1) tells us that there are three competing terms which
drive the expansion: a term of energy (matter and radiation), the
cosmological constant, and a  curvature term. Recent observations
on temperature anisotropy of the cosmic microwave
background~\cite{Balbi:2000tg,deBernardis:2001xk,Bennett:2003bz}
strongly indicate that the Universe geometry is very close to
flat. So that, in the following we  assume $k \, = \, 0$. We are
interested in the matter-dominated Universe where the radiation
energy density and the matter pressure are negligible. Allowing
for a FDW gas with density $\rho_{w}$ and pressure $p_{w}$ related
by the equation of state Eq.~(4.3), the Friedmann-Lemaitre
equations become:
\begin{equation}
H^2 \; = \; \frac{8\pi G_N}{3}\,(\rho_{m} \, + \, \rho_{w})  \, +
\, \frac{\Lambda}{3},
\end{equation}
\begin{equation}
2 \frac{\stackrel{\cdot \cdot}{R}}{R} + \; H^2 \; = \; -8\pi G_N
\,p_{w} \; + \; \Lambda \; ,
\end{equation}
where
\begin{equation}
H \; = \;  \frac{\stackrel{\cdot}{R}}{R}
\end{equation}
is the Hubble parameter.
By introducing the following $\Omega$-functions:
\begin{eqnarray}
\Omega_{m} &\!\! \equiv \!\!&  \frac{\rho_{m}}{  \rho_{c}} \, , \\
\Omega_{w} &\!\! \equiv \!\!& \frac{ \rho_{w}}{  \rho_{c}} \, , \\
\Omega_{\Lambda} &\!\! \equiv \!\!&  \frac{ \Lambda}{ 3 H^2},
\end{eqnarray}
where $\rho_{c} \equiv 3H^2/8\pi G_N$ is the critical density, we
cast Eq.~(5.3) in the form:
\begin{equation}
\Omega_{m} + \Omega_{w} + \Omega_{\Lambda} = 1.
\end{equation}
By defining the following dimensionless variables:
\begin{eqnarray}
\tau &\!\! \equiv  \!\!& t/t_{0} \, , \\
r &\!\! \equiv  \!\!&  R/R_{0} \, ,
\end{eqnarray}
where the index $0$ refers to the present time, and using
Eq.~(4.3), the equation of state for the ideal gas of FDW, we
write Eq. (5.4) in the form:
\begin{equation}
\left(\frac{dr}{d\tau}\right) \; = \; t_0  H_0 \, \left[
\Omega_{m0} \, r^{-1} \, + \, \Omega_{w0} \, r^{-2} \, + \,
\Omega_{\Lambda0} \, r^{2}\right]^{1/2} \; .
\end{equation}
In Figure~1 we compare the expansion parameter as a function of
time for three different models. We have fixed $\Omega_{m0}=0.3$
and consider ($\Omega_{\Lambda0}=0.0$ , $\Omega_{w}=0.7$) ,
($\Omega_{\Lambda0}=0.7$ , $\Omega_{w0}=0.0$) and
($\Omega_{\Lambda0}=0.6$ , $\Omega_{w0}=0.1$). In the first case
$R$ behaves quite similar to the Einstein-de~Sitter model. The
standard cosmological  model corresponds to $\Omega_{m0}=0.3$,
$\Omega_{\Lambda0}=0.7$. From Fig.~1 we see that the three models
are almost identical up to the present epoch ( $t \, H_0 \,
\lesssim \, 1.7$). To discriminate among models we compare the
Friedman-Robertson-Walker magnitude-redshift relation with the
combined low and high redshift supernova dataset. To this end, we
recall that the luminosity-distance is:
\begin{equation}
D_{L}(z) = \left(\frac{{\cal L}}{4\pi {\cal F}}\right)^{\!\!1/2}
\!\!\!\!\!\!,
\end{equation}
where $z$ is the redshift relative to the present epoch, ${\cal
L}$ is the intrinsic luminosity of the source, and ${\cal F}$ the
observed flux. Setting $d_{L}(z) \equiv H_{0}D_{L}(z)$, for
Friedmann-Lemaitre models it is well known  that:
\begin{eqnarray}
d_{L}(z) &\!\! =  \!\!& (1+z)\frac{1}{\sqrt{ -\Omega_{k0}}}\sin
                  \left( \sqrt{ -\Omega_{k0}}\, \int_{0}^{z}\!\frac
                  {dz'}{H(z')/H_{0}}
                  \right)\;\;\; \mbox{if $\Omega_{k0} < 0$}, \\
d_{L}(z) &\!\! = \!\!&  (1+z)\int_{0}^{z}\!\frac
                  {dz'}{H(z')/H_{0}}
                \;\;\;\;\;\;\;\;\;\;\;\;\;\;\;\;\;\;\;\:\;\;\;\;\;\;\;\;\;\;\;\;\;\;\;\;\;\;\;
                   \mbox{if $\Omega_{k0} = 0$}, \\
d_{L}(z) &\!\! = \!\!&  (1+z)\frac{1}{\sqrt{ \Omega_{k0}}}\sinh
                 \left( \sqrt{ \Omega_{k0}}\, \int_{0}^{z}\!\frac {dz'}{H(z')/H_{0}}
                 \right)\;\;\;\;\;\;\, \mbox{if $\Omega_{k0} > 0$},
\end{eqnarray}
where the Hubble parameter is given by
\begin{equation}
H(z) = H_{0}\: \left[ \,\Omega_{m0}(1+z)^3 \; + \;
\Omega_{\Lambda0} \; + \; \Omega_{w0}(1+z)^4 \, \right]^{1/2} \, .
\end{equation}
The apparent magnitude $m$ of an object is related to its redshift
by the following relation:
\begin{equation}
m(z) = 5\log_{10}d_{L}(z)+ \cal{M},
\end{equation}
where ${\cal M} = M + 5\log_{10}(\mbox{Mpc}/D_{H}) + 25$ is the
Hubble-constant-free absolute magnitude, $M$ is the absolute
magnitude, and $D_{H} \equiv c/H_{0}$ is the Hubble distance. \\
Specializing to a flat Universe in Fig.~2 we display $m(z)$ in
Eq.~(5.18) as a function of the redshift $z$ for the three models.
The recent SN Ia sample provides measurements of the luminosity
distance out to redshift $ z \sim 1.7 $ (SN 1997ff).
In Figure~2 we also shows the Hubble diagram of effective
rest-frame B magnitude corrected for the width-luminosity
relation, as a function of redshift for the 42 Supernova Cosmology
Project high-redshift supernovae~\cite{Perlmutter:1998np}, along
with the 18 Calan/Tololo low-redshift
supernovae~\cite{Hamuy:1996}. We  also display the 10 High-Z
Supernova search Team supernovae with $ 0.16 \leq z \leq
0.62$~\cite{Riess:1998}, and the recent SN 1997ff at $ z \sim
1.7$~\cite{Riess:2001gk}. In the case of the farthest known
supernova SN 1997ff we also report the revised value of effective
rest-frame B magnitude corrected  after correction for
lensing~\cite{Benitez:2002jk}.
\\
Even though we do not attempt to best fit  the supernova data,
from Fig.~2 it is evident  that the model with $\Omega_{m0}=0.3$,
$\Omega_{\Lambda0}=0.0$ , $\Omega_{w}=0.7$ is excluded. On the
other hand, both the standard model and the model with
$\Omega_{m0}=0.3$, $\Omega_{\Lambda0}=0.6$ , $\Omega_{w}=0.1$ are
consistent with the high-z supernova data.
\footnote{It is worthwhile to stress that the perfect gas
approximation (which implies that a network of FDWs behaves as a
radiation gas) cannot be valid when interactions with the plasma
are taken into account. Indeed, if the equation of state were
valid at all times then, due to the large value of relativistic
matter ($\Omega_w = 0.1$), the equilibrium between matter and
radiation would take place later respect to the cosmological
standard model and the structure formation would proceed at too
small $z$. It is clear that this would be in contrast with the
standard analysis of structure formation (see e.g. \cite{KOLB}).
On the other hand, assuming an equation of state more stiff for
very high redshifts, such high values for $\Omega_w$ are not a
priori excluded.}
This can be better appreciated from Fig.~3 where we display the
difference between data and the fiducial standard model
$\Omega_{m0}=0.3$, $\Omega_{\Lambda0}=0.7$, $\Omega_{w}=0.0$. Thus
we see that recent high-z supernova data allow that a small, but
cosmological important, part of our Universe consists of an almost
ideal gas of ferromagnetic domain walls.
\renewcommand{\thesection}{\normalsize{\arabic{section}.}}
\section{\normalsize{CMBR constraints}}
\renewcommand{\thesection}{\arabic{section}}
As discussed in Section~I, in general we expect that a network of
domain wall extending over cosmological distance could lead to
severe distortions of the Cosmic Microwave Background Radiation.
Thus, it is important to investigate the influence of the
cosmological gas of massive ferromagnetic domain walls  on the
Cosmic Microwave Background Radiation. If domain walls are present
we have an additional temperature anisotropy:
\begin{equation}
\frac {\delta T} {\langle T \rangle } \; =  \; \Delta
\Phi \; ,
\end{equation}
where $\Phi(r)$ is the Newtonian potential of the wall network. To
evaluate the temperature anisotropy Eq.~(6.1) we follow the
analysis performed in Ref.~\cite{FRIEDLAND}. We find
\begin{equation}
\frac {\delta T} {\langle T \rangle } \; = \; \Phi(\langle L
\rangle ) \, n^{\nu} \; ,
\end{equation}
where $\langle L \rangle$ is the average distance between walls
and $n$ is the number of walls per horizon. As concern the
exponent $\nu$, it turns out that this exponent depends on the
network walls configuration. According to Ref.~\cite{FRIEDLAND} we
consider values of $\nu$ between  $\nu \, = \, 1$ (regular
network) and $\nu \, = \, 3/2$ (non-evolving network).
In our case, the Newtonian gravitational potential for a FDW in
the $(x,y)$ plane is $\phi(z)=4 \pi G_N |z| \sigma_w$
\footnote{Given the stress tensor
$T^{\mu}_{\nu}=diag(\rho,-p_1,-p_2,-p_3)$, the Newtonian limit of
Poisson's equation is $\nabla^{2} \phi=4 \pi G_N (\rho
+p_1+p_2+p_3)$, where $\phi$ is the Newtonian gravitational
potential. For a thin FDW $\rho=\delta(z) \sigma_w$,
$p_1=p_2=\rho/2$ and $p_3=0$. Thus, $\nabla^{2} \phi=8 \pi G_N
\delta(z) \sigma_w$.}.
Moreover we have $n= d_H / \langle L \rangle$, so that from
Eq.~(\ref{eq4.2}) we obtain:
\begin{equation}
\frac {\delta T} {\langle T \rangle } =  4 \, \pi \, G_N \,
\sigma_w^{2- \nu} \gamma^{2(1- \nu)} \, d_H^{\nu} \,
\Omega_{w}^{\nu-1} \, \rho_c^{\nu-1}.
\end{equation}
The quantities in Eq.~(6.3) need to be evaluated  at the present
time $t_0$. Let us observe that:
\begin{equation}
d_H(t_0) = R(t_0) \int^{t_0}_0 \!\! \frac {d t} {R(t)} = t_0
\int^1_0 \!\! \frac {d \tau} {r(\tau)} \, ,
\end{equation}
where $r(\tau)= R(t)/R(t_0)$, $\tau=t/t_0$,  and
\footnote{From  energy conservation we have that $\rho_w \sim
R^{-4}$; since $\rho_w \sim \sigma_w/\langle L \rangle$ and $
\langle  L \rangle \sim R$, we get $\sigma_w \sim R^{-3}$.}
\begin{equation}
\sigma_w(t_0) = \sigma_w(t_{ew}) \left( \frac {R(t_{ew})} {R(t_0)}
\right)^{\!\!3} = \sigma_w(t_{ew}) \, r_{ew}^3 \, ,
\end{equation}
 $t_{ew}$ denoting the time at the electroweak phase transition. \\
Now, we recall that $r(\tau)$ is solution of the equation:
\begin{equation}
\label{CMB6}
\frac {d r (\tau)} {d \tau}= t_0 H_0 \, \left[ \Omega_{m0} r^{-1}
+ \Omega_{w0} r^{-2} + \Omega_{\Lambda0} r^2 \right]^{1/2}.
\end{equation}
Putting $f(r)=(\Omega_{m0} r^{-1} + \Omega_{w0} r^{-2} +
\Omega_{\Lambda0} r^2)^{1/2}$ we have
\begin{equation}
\label{CMB7}
t_0= H_0^{-1} \int_0^1 \!\! \frac{d r}{f(r)} \, .
\end{equation}
So that we can write:
\begin{equation}
\label{CMB9}
\frac {\delta T} {\langle T \rangle } = \alpha_{\nu} \, [\,  G_N
\, \sigma_w(t_{ew}) \, t_{ew}^{3/2} \, H_0^{1/2}\, ]^{\,2- \nu} ,
\end{equation}
where
\begin{equation}
\label{CMB10}
\alpha_{\nu} = \frac {256} {3} \pi^2 \left(\frac {3 \sqrt{2}} {32
\pi} \right)^{\!\nu} \gamma^{2(1- \nu)} \, \Omega_w^{\frac {\nu+2}
{4}} \left(\int_0^1 \!\! \frac {d x} {r(x)} \,  \int_0^1 \!\!
\frac {d x} {f(x)} \right)^{\!\nu} \!\! .
\end{equation}
Eqs.~(\ref{CMB9},\ref{CMB10}) give an approximate estimate of
$\,\delta T/ \langle T \rangle $. \\
Taking $ G_N \, \sim(10^{19}\, \text{GeV})^{-2}$,
$\sigma_w(t_{ew}) \, \sim \, 10^6 \, \text{GeV}^3$, $t_{ew}\, \sim
\, 10^{13}\, \text{GeV}^{-1}$, $H_0 \, \sim \, 10^{-42}\text{
GeV}$~\cite{KOLB}, and assuming $\Omega_m=0.3$, $\Omega_w=0.1$,
$\Omega_{\Lambda}=0.6$, we obtain:
\begin{eqnarray}
\label{CMB11}
\frac {\delta T} {\langle T \rangle } \!\! & \sim & \!\! 10^{-33}
\;\;\;\; \text{for} \,\,\, \nu=1 \, , \\ \nonumber \frac {\delta
T} {\langle T \rangle }  \!\! & \sim & \!\! 10^{-17} \;
\gamma^{-1}  \;\;\;\; \text{for} \,\,\, \nu= 3/2\, .
\end{eqnarray}
Even in the worst case $\nu \, = \, 3/2$ the contributions to the
large-scale anisotropy of the microwave background radiation is
completely negligible given the observed value $\delta T / T \sim
10^{-6}$~\cite{Balbi:2000tg,deBernardis:2001xk,Bennett:2003bz}.
%
%
%
%********************************************************************************************%
%
%
%
\renewcommand{\thesection}{\normalsize{\arabic{section}.}}
\section{\normalsize{Conclusions}}
\renewcommand{\thesection}{\arabic{section}}
Let us conclude by stressing the main results of the paper. From
the constraint that domain walls must not destroy the observed
isotropy of the Universe, it is widely believed that domain walls
that might have been produced in the standard particle physics
phase transitions in the very early Universe have been eliminated
by some mechanism. We have discussed a new mechanism which allows
domain walls produced during the primordial electroweak phase
transition. Indeed, the effective surface tension of these domain
walls can be made vanishingly small due to a peculiar magnetic
condensation induced by fermion zero modes localized on the wall.
As a consequence, the domain wall acquires a non zero magnetic
field perpendicular to the wall, and it becomes almost invisible
as far as gravitational effects are concerned. We find that in the
perfect gas approximation the domain wall network behaves like a
radiation gas. The analysis of the recent high-redshift supernova
data suggests that a small component of our Universe could be
composed of this gas of ferromagnetic domain walls. Finally, we
showed that ferromagnetic domain wall gas induces a completely
negligible contribution to the large-scale anisotropy of the
microwave background radiation.

Undoubtedly, the presence of a network of massive FDWs can have
influenced physical processes that took place in the early
Universe. It should keep in mind that in studying effects occurred
in the early Universe (say before or during the formation of large
scale structures) the perfect gas approximation in no longer valid
and a more complicated analysis of the interaction of walls with
the primordial plasma must be in order. We reserve such subject
matter to future investigations.
\newpage
\appendix
\section{\normalsize{Appendix}}
We are interested in the Dirac equation with a  kink wall and in
presence of the electromagnetic field $ A_{\mu}(x)$:
\begin{equation}
(i \, \slash\!\!\!\partial - e \; \slash\!\!\!\!\!\:\!A - g_{Y} \,
\phi ) \, \psi = 0 \; .
\end{equation}
To solve Eq.~(A.1) we assume that
\begin{equation}
\psi = (i \, \slash\!\!\!\partial - e \; \slash\!\!\!\!\!\:\!A +
g_{Y} \, \phi ) \, \Phi.
\end{equation}
Inserting Eq.(A.2) into Eq.~(A.1) we readily obtain:
\begin{equation}
(- \partial^{2} - 2ieA_{\mu} \partial^{\mu} + e^2 A^2 - \xi^2 g^2
+ i \xi \, \slash\!\!\!\partial g - \frac{e}{2} \sigma^{\mu \nu}
F_{\mu \nu}) \, \Phi = 0,
\end{equation}
with $\xi = g_{Y} \, v $, $g(z)=\tanh (z / \Delta)$. It is easy to
see that the solution of Eq.~(A.3) factorize as:
\begin{equation}
\Phi(x,y,z,t) = f(x,y) \, \chi(z,t),
\end{equation}
where $f$ is a scalar function which describes the motion
transverse to the wall and $\chi$ is the spinorial part of the
wave function of fermions localized on the wall. Putting $
\chi(z,t) = e^{-iEt}\, \chi(z)$ we get for $f$ and $\chi$:
\begin{equation}
(\partial_{x}^{2} + \partial_{y}^{2} - 2ieBx \partial_{y} - e^2
B^2 x^2 + \beta) \, f(x,y) = 0,
\end{equation}
\begin{equation}
(\partial_{z}^{2} + i \xi \gamma^3 \partial_{z} g  - \xi^2 g^2 +
ieB \gamma^1 \gamma^2 + \alpha) \, \chi(z) = 0,
\end{equation}
where $\alpha$ and $\beta$ are constants subject to following
relation:
\begin{equation}
E^2 = \alpha + \beta.
\end{equation}
The solution of Eq.~(A.5) is:
\begin{equation}
f_{n,p}(x,y) = A_{n} \, e^{-\frac{1}{2} \zeta^{2} \,+ \, ip y} \,
H_{n}(\zeta),
\end{equation}
where
\begin{equation}
A_n = \frac{\pi^{-1/4}}{\sqrt{2^{n}n!}} \, , \;\;\;\; \zeta =
\sqrt{|e|B} \left(x-\frac{p}{eB}  \right),
\end{equation}
$H_{n}(x)$ are Hermite polynomials and
\begin{equation}
\label{eqA.10} \beta = |e|B (2n+1), \;\;\;\;\; n=0,1,2,...
\end{equation}
In order to solve Eq.~(A.6), we expand $\chi(z)$ in terms of
spinors $u_{\pm}^{1,2}$ eingenstates of $\gamma^3$
\begin{equation}
\chi(z) = \phi_{\pm}^{1,2} u_{\pm}^{1,2}.
\end{equation}
Using the standard representation for the Dirac
matrices~\cite{BJORKEN} we find:
\begin{equation}
u_{\pm}^{1} = \left( \begin{array}{c}
                     1 \\
                     0 \\
                     \pm i \\
                     0
                     \end{array} \right) \! ,
           \;\;\;\;\; u_{\pm}^{2} = \left( \begin{array}{c}
                     0 \\
                     1 \\
                     0 \\
                     \mp i
                     \end{array} \right) \! .
\end{equation}
It is straightforward to check that:
\begin{eqnarray}
\gamma^0 u_{\pm}^{1,2} \!\!&=&\!\! u_{\mp}^{1,2}, \\ \nonumber
\gamma^1 u_{\pm}^{1} \!\!&=&\!\! \pm i u_{\mp}^{2} \, ,\\
\nonumber
\gamma^1 u_{\pm}^{2} \!\!&=&\!\! \mp i u_{\mp}^{1} \, ,\\
\nonumber
\gamma^2 u_{\pm}^{1,2} \!\!&=&\!\! \mp u_{\mp}^{2,1}, \\
\nonumber \gamma^3 u_{\pm}^{1,2} \!\!&=&\!\! \pm i u_{\pm}^{1,2}.
\end{eqnarray}
Taking into account Eqs.~(A.6), (A.11) and (A.13) we obtain:
\begin{equation}
(\partial_{z}^{2} \mp \xi \partial_{z} g  - \xi^2 g^2 + eB +
\alpha) \, \phi_{\pm}^{1}(z) = 0,
\end{equation}
\begin{equation}
(\partial_{z}^{2} \mp \xi \partial_{z} g  - \xi^2 g^2 - eB +
\alpha) \, \phi_{\pm}^{2}(z) = 0.
\end{equation}
It easy to see that there are localized states if
\begin{eqnarray}
\alpha \!\!&=&\!\! -eB \;\;\;\; \mbox{for} \;\; \phi_{\pm}^{1}, \\
\alpha \!\!&=&\!\! +eB \;\;\;\; \mbox{for} \;\; \phi_{\pm}^{2}.
\end{eqnarray}
Inserting Eqs.~(A.10), (A.16), (A.17) into Eq.~(A.7) we get the
energy spectrum for localized states:
\begin{equation}
E_n = \sqrt{2n|e|B} \;\;\;\;\; n=1,2,...
\end{equation}
We can now  solve Eqs.~(A.14) and (A.15) whit the constrains
(A.16) and (A.17); we find:
\begin{equation}
\phi_{\pm}^{1,2}(z) = N \, \exp{\left(\pm \, \xi \int_{0}^{z} \!
dz' g(z') \right)},
\end{equation}
where  the normalization constant $N$ is  evaluated below. The
solutions $\phi_{+}^{1,2}$  do not correspond to  localized states
and will be discarded. Inserting Eqs.~(A.19), (A.11), (A.8), and
(A.4) into Eq.~(A.2) it easy to recover our Eqs.~(3.2)-(3.5).
Finally,  the normalization constant $N$ can be obtained by
imposing the normalization conditions:
\begin{equation}
\int{ \! d^3x \, \psi^{\dagger}_{n,p,s}({\bf x}) \,
\psi_{n',p',s'} ({\bf x})} = \delta_{n n'} \, \delta_{s s'} \,
\delta (p-p')\; .
\end{equation}
It is straightforward to obtain:
\begin{equation}
N^2 = \frac{\sqrt{|e|B}}{8\pi \Delta \mbox{B}(\xi \Delta,1/2) \,
E^2} \, \; .
\end{equation}
\vfill
\newpage
\renewcommand{\thesection}{\normalsize{\arabic{section}.}}

%
%
%
%********************************************************************************************%
%
%
%
\vfill
\newpage
%
%
%
%********************************************************************************************%
%
%
%
\begin{figure}[H]
\label{Fig1}
\begin{center}
\includegraphics[clip,width=1.05\textwidth]{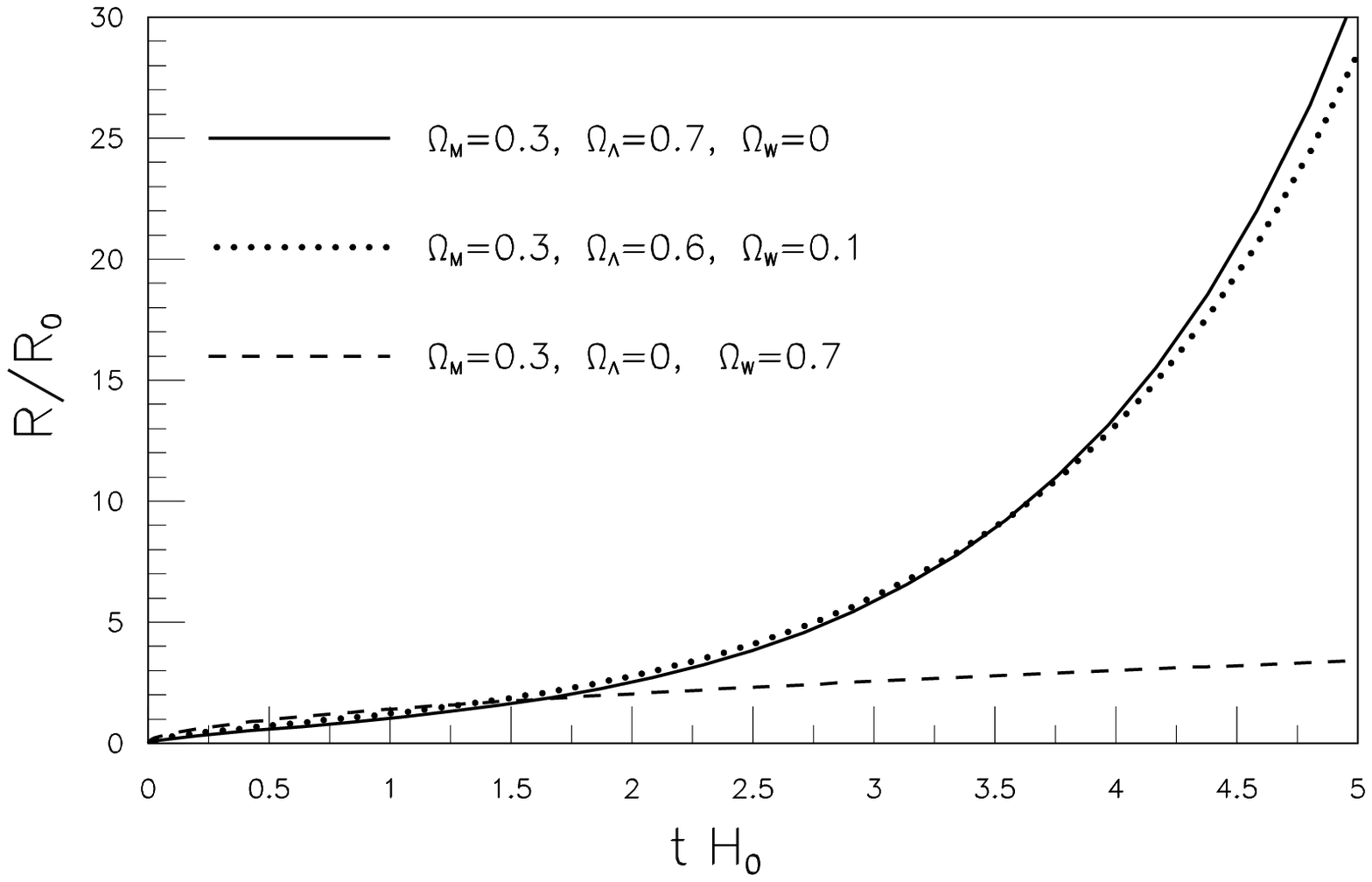}
\caption{Evolution of expansion parameter for three different
models.}
\end{center}
\end{figure}
%
%
%
%********************************************************************************************%
%
%
%
\begin{figure}[H]
\label{Fig2}
\begin{center}
\includegraphics[clip,width=1.05\textwidth]{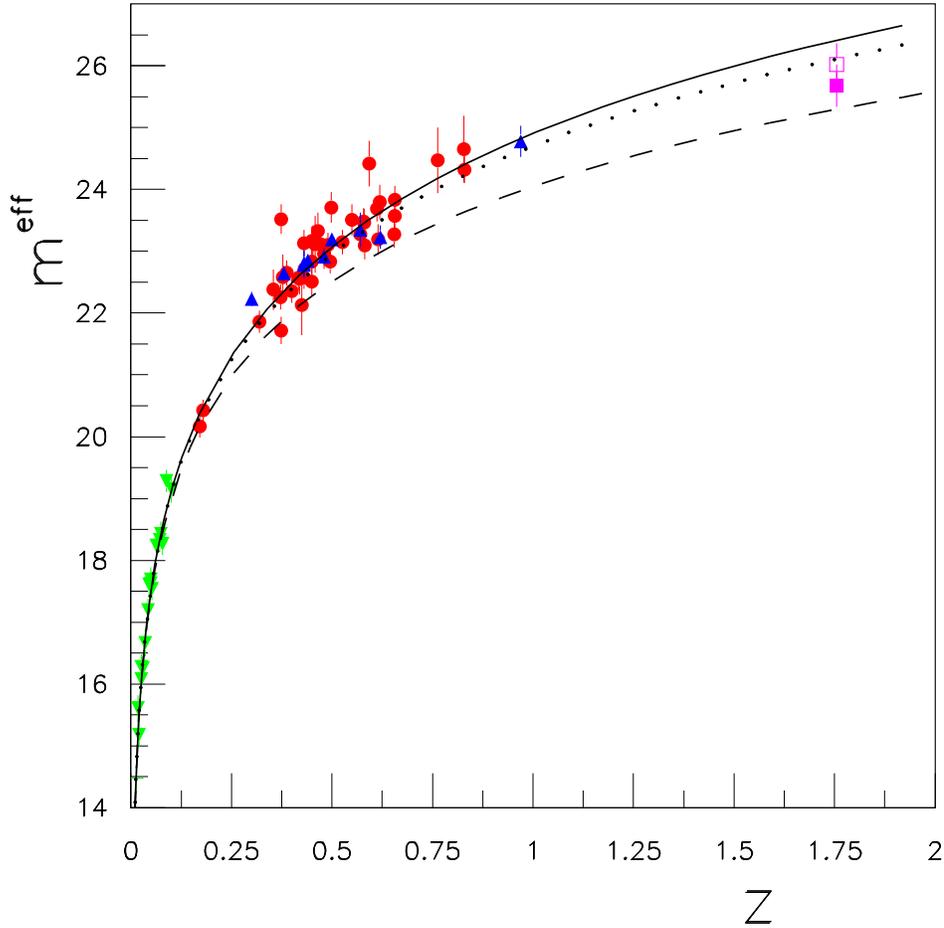}
\caption{Hubble diagram for Type Ia supernovae data.  The meaning
of symbols is as follows. Calan/Tololo supernovae (full triangles
down), Supernova Cosmology Project high-redshift supernovae (full
circles), High-Z Supernova search Team supernovae ( full triangles
up), SN 1997ff (full square) and SN 1997ff after correction for
lensing (open square). Full line corresponds to the standard model
$\Omega_m=0.3$, $\Omega_{\Lambda}=0.7$ and $\Omega_w=0$, dashed
line to $\Omega_{m0}=0.3$, $\Omega_{\Lambda0}=0.0$ ,
$\Omega_{w}=0.7$, dotted line to $\Omega_{m0}=0.3$,
$\Omega_{\Lambda0}=0.6$ , $\Omega_{w}=0.1$.}
\end{center}
\end{figure}
%
%
%
%********************************************************************************************%
%
%
%
\begin{figure}[H]
\label{Fig3}
\begin{center}
\includegraphics[clip,width=1.05\textwidth]{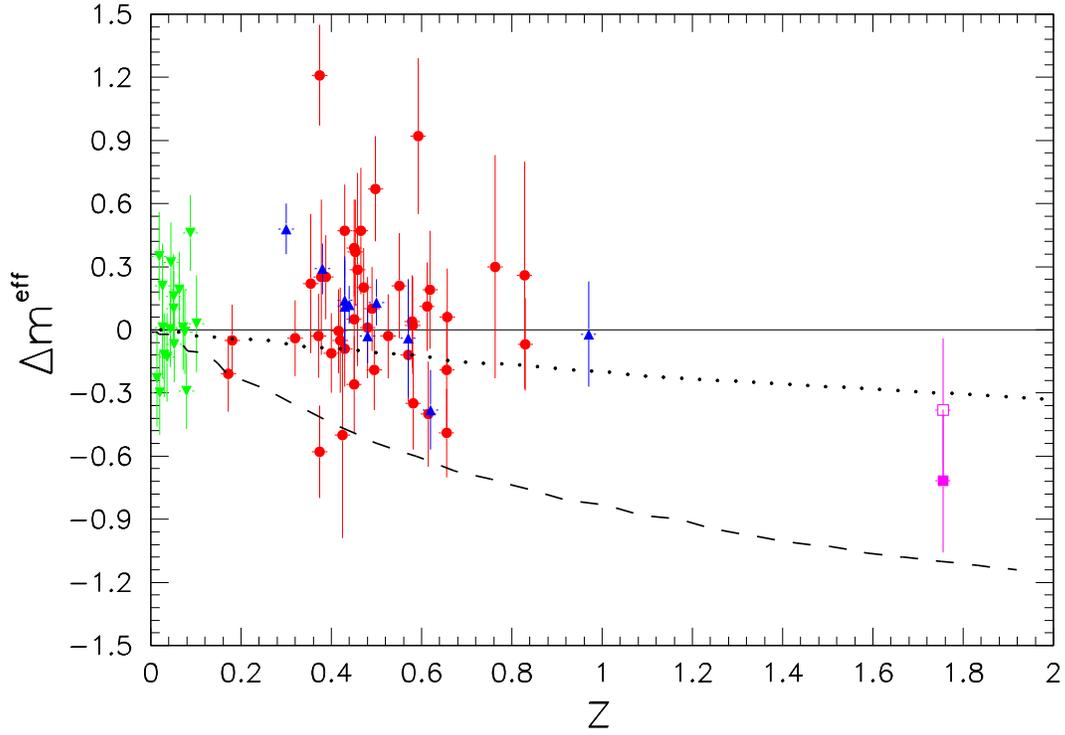}
\caption{The magnitude residuals between data and models
$\Omega_{m0}=0.3$, $\Omega_{\Lambda0}=0.0$ , $\Omega_{w}=0.7$
(dashed line),  $\Omega_{m0}=0.3$, $\Omega_{\Lambda0}=0.6$ ,
$\Omega_{w}=0.1$ (dotted line) from the standard model
$\Omega_m=0.3$, $\Omega_{\Lambda}=0.7$ and $\Omega_w=0$ (full
line).}
\end{center}
\end{figure}
\end{document}